\documentclass[sn-mathphys,Numbered,article]{sn-jnl}
 \usepackage{graphicx}%
\usepackage{multirow}%
\usepackage{amsmath,amssymb,amsfonts}%
\usepackage{amsthm}%
\usepackage{mathrsfs}%
\usepackage[title]{appendix}%
\usepackage{xcolor}%
\usepackage{textcomp}%
\usepackage{manyfoot}%
\usepackage{booktabs}%
\usepackage{algorithm}%
\usepackage{algorithmicx}%
\usepackage{algpseudocode}%
\usepackage{listings}%
\usepackage[sorting=none,maxnames=999]{biblatex}%
\theoremstyle{thmstyleone}%

\theoremstyle{thmstyletwo}%

\theoremstyle{thmstylethree}%

\begin{document}

\title[Article Title]{QEAN: Quaternion-Enhanced Attention Network for Visual Dance Generation}


\author[1]{Zhizhen Zhou}
\email{3121005415@mail2.gdut.edu.cn}
\author[1]{Yejing Huo}
\email{2112305121@mail2.gdut.edu.cn}
\author*[1]{Guoheng Huang}
\email{kevinwong@gdut.edu.cn}
\author[1]{An Zeng}
\email{zengan@gdut.edu.cn}
\author*[2]{Xuhang Chen}
\email{xuhangc@hzu.edu.cn}
\author[3]{Lian Huang}
\email{mrhuanglian@gmail.com}
\author[4]{Zinuo Li}
\email{zinuo.li@research.uwa.edu.au}
\affil[1]{Guangdong University of Technology, Guangdong, China}
\affil[2]{Huizhou University, Guangdong, China}
\affil[3]{Guangdong Mechanical and Electrical College, Guangdong, China}
\affil[4]{University of Western Australia, WA, Australia}

\abstract{
The study of music-generated dance is a novel and challenging Image generation task. It aims to input a piece of music and seed motions, then generate natural dance movements for the subsequent music. Transformer-based methods face challenges in time series prediction tasks related to human movements and music due to their struggle in capturing the nonlinear relationship and temporal aspects. This can lead to issues like joint deformation, role deviation, floating, and inconsistencies in dance movements generated in response to the music. In this paper, we propose a Quaternion-Enhanced Attention Network (QEAN) for visual dance synthesis from a quaternion perspective, which consists of a Spin Position Embedding (SPE) module and a Quaternion Rotary Attention (QRA) module. First, SPE embeds position information into self-attention in a rotational manner, leading to better learning of features of movement sequences and audio sequences, and improved understanding of the connection between music and dance. Second, QRA represents and fuses 3D motion features and audio features in the form of a series of quaternions, enabling the model to better learn the temporal coordination of music and dance under the complex temporal cycle conditions of dance generation. Finally, we conducted experiments on the dataset AIST++, and the results show that our approach achieves better and more robust performance in generating accurate, high-quality dance movements. Our source code and dataset can be available from \url{https://github.com/MarasyZZ/QEAN} and \url{https://google.github.io/aistplusplus_dataset} respectively.
}

\keywords{Dance generation, Multi-modal task, Quaternion network, Time-series prediction task, Animation generation task}

\maketitle

\section{Introduction}\indent

\begin{figure*}[htbp]
  \centering
  \includegraphics[width=1.0 \linewidth]{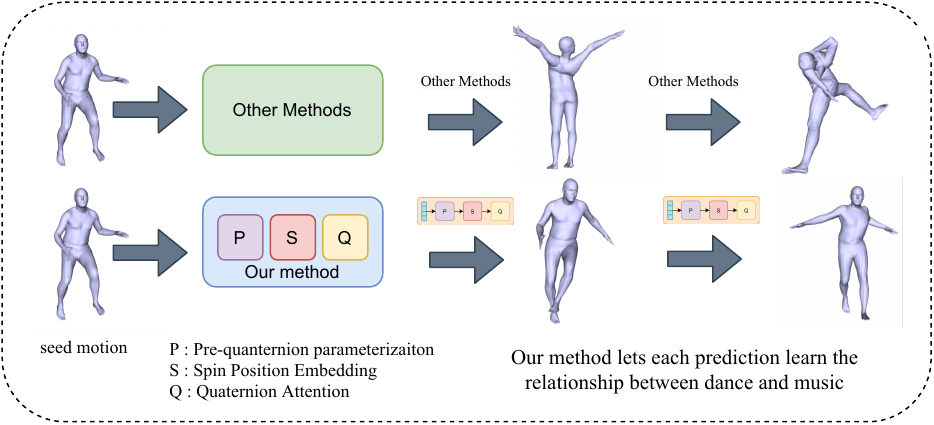} 
  \caption{
        The motivation of our method. We compare the effectiveness of our method is compared with other approaches in generating dance movements from seed motions.  In the top row labeled ``other methods", two sets of images  showcase the transformation of seed movements into unnatural final poses characterized by joint deformation and character drift. Conversely, in the bottom row labeled ``our method", we demonstrate how the application of Pre-quaternion parameterization (P), Spin Position Embedding (S), and Quaternion Attention (Q) yields natural-looking final poses. Each prediction produced by our method successfully learns the correlations between dance and music.
  }
  \label{fig:1}
\end{figure*}

Dancing is a universal language across all cultures \cite{bib1, bib2} and is used by many as a powerful means of self-expression on online media platforms, becoming a dynamic tool for disseminating information on the Internet. Although dance is an art form, it requires professional practice and training to give dancers a rich expressive voice \cite{bib66}. Therefore, from a computational point of view, music-conditioned 3D dance generation \cite{bib3,bib4,bib5,bib6} has become a critical task that promises to open up a variety of practical applications. However, creating satisfying dance sequences that harmonize with specific music and body structures faces challenges because of our lack of understanding of the timing of human movements and the connection between music and dance. Overcoming these challenges is essential to achieve fluid movements with a high degree of kinematic complexity while ensuring consistency with the complex non-linear relationships of the accompanying music.

Later, with the continuous development and advancement of deep learning, many deep learning methods \cite{bib5,bib6,bib25,bib69,bib58,bib67} were started to be applied to dance generation. Firstly, RNN \cite{bib25,bib58} based methods were used to simulate human dances, but RNN approaches would face the challenges of static poses and error accumulation, especially when the input data varied. Subsequently, some researchers have used Variational Auto-Encoders (VAE) and Generative Adversarial Networks (GAN) to model 2D dance movements \cite{bib61}, and then LSTM-Auto-Encoders were used to model 3D dance movements directly from musical features \cite{bib62}, although such an approach solves the shortcomings of error accumulation that exist in the RNN approach. However, such an approach suffers from the disadvantage of instability and is prone to regress to non-standard poses. 
 
In recent years, Transformer \cite{bib61,bib62,bib22,bib68, Li_2023_ICCV, ijcai2023p129, luo2023devignet} has been favoured by many in natural language processing as well as visual processing, and some scholars have made great progress in their research \cite{bib7,bib34} to be able to generate high-quality dance movements given a piece of music. However, due to the existence of Transformer's inadequate modeling of the temporal dependence of sequences when dealing with time-series data, the generated dances will suffer from problems such as drifting and foot slipping (As shown in Fig.\ref{fig:1}). Given the non-linear relationship between music and dance, existing approaches to Transformer do not fully model this relationship. 

Quaternions are widely used as a mathematical tool for rotational expression and gesture control \cite{bib64}. In view of this, we believe that the introduction of quaternions in the field of dance generation may be a promising approach. Compared to traditional Euler angles, quaternions are more effective in avoiding the ``Gimbal Lock" problem and improving the stability of gesture representations. We expect to use quaternions to more accurately adapt dance movements to the rhythm and emotion of the music. By combining musical features with the correlation of quaternions, we can more accurately capture the complex relationship between music and dance.

\begin{figure*}[htbp]
  \centering
  \includegraphics[width=1.0 \linewidth]{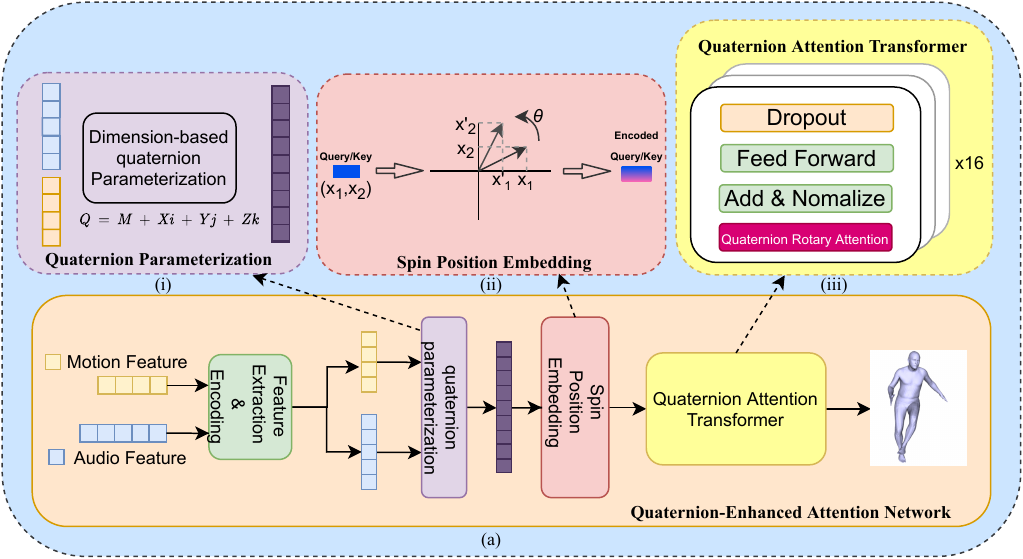}
  \caption{The overview of our method. (a) describes the basic process, which contains three modules (i), (ii), and (iii). When the inputs are a motion sequence with a length of 120 frames and an audio sequence with a length of 240 frames, features are extracted by the motion transformer and the audio transformer, respectively. The extracted features are parameterized by a quadratic parameterization operation, and the dimension is changed to 4 dimensions. Through the Spin Position Embedding (SPE) module, the corresponding 4-dimensional features are rotated to embed the information into the self-attention in a rotational manner. The information processed by the SPE is used to explore the coordination between the music and the dance through the quaternionic attentional transformer, and finally, the corresponding dance is generated. (i), (ii) and (iii) describe the processing of quaternion parameterization, spin position embedding and the basic structure of the quaternion attention transformer, respectively. Specific details are given in the Methods section.}
  \label{fig:2}
\end{figure*}

In this paper, to address these challenges, we propose a Quaternion-Enhanced Attention Network for multi-modal dance synthesis (shown in Fig.\ref{fig:2}). The network mainly consists of a Spin Position Embedding (SPE) module and a Quaternion Rotary Attention (QRA) module. The SPE module is mainly used in the Transformer structure of the network, which embeds information in the form of relative positions into the self-attention. The audio and motion features are extracted by the Transformer structure in the network, respectively, and the SPE module combines the advantages of relative position coding and absolute position coding to maximize the model's representation of sequence features. The extracted audio and motion features are expanded to four dimensions by quaternion parameterization dimension, and then the splicing operation is performed through the Quaternion Rotary Attention module. Compared to the work of Li \cite{bib7}, the proposed SPE better increases the model's representation and utilisation of positional information. Besides, the QRA module better learns the representation of the link between audio and movement relationships, improving the quality of the generated dances with good robustness.

The contribution of the proposed network can be summarized as follows:

1. In this paper, we introduce a Quaternion-Enhanced Attention Network (QEAN) for generating multimodal dances. This addresses challenges seen in current methods, like awkward joint movements and character inconsistency. QEAN uses quaternion operations to better capture the complex relationship between music and dance, improving the modeling of temporal dependencies.

2. We introduce the Spin Position Embedding (SPE) module, which computes query and key vectors for features, applies rotational operations, and embeds results into self-attention. SPE addresses limitations of traditional position encoding by introducing relative position encoding based on rotations, enhancing modeling for variable-length sequences while solving length consistency and overfitting issues. Additionally, relative position information improves modeling of intrinsic feature associations, significantly enhancing the model's representation and utilization of temporal order in human motion.

3. We introduce the quaternion perspective and propose the Quaternion Rotary Attention (QRA) module. The QRA module maps audio and motion features to the quaternion space and explores the intrinsic correlation between the two using Hamiltonian multiplication, which enables the model to better learn the temporal coordination between music and dance, and generate smooth and natural dances coordinated with the music tempo. 

4. Experimental results on the AIST++ dataset demonstrate that our proposed network is capable of effectively learning the connection between audio and movement, leading to the generation of  higher quality dance movements. It outperforms other current state-of-the-art methods in terms of dance quality.

\section{Related Work}\indent

\textbf{3D Human Motion Synthesis} The research on generating realistic and controllable 3D human motion sequences, as discussed in \cite{bib25,bib12,bib18, bib35}, has seen significant advancements in recent years. Initial efforts utilized statistical models like kernel-based probability distributions \cite{bib39} to synthesize motion, but these methods tended to oversimplify motion details. A subsequent breakthrough came with the introduction of the motion graph approach \cite{bib40}, which addressed this limitation by adopting a non-parametric method. This technique involved constructing a directed graph using motion capture datasets, where each node represented a pose, and edges denoted transitions between poses. Motion generation was achieved through random walks on this graph. However, a notable challenge in motion graphs was the generation of plausible transitions, and certain methods sought to overcome this by introducing parameterizations for transitions \cite{bib42}. As deep learning gained prominence, several approaches explored the use of neural networks trained on extensive motion capture datasets to generate 3D motion. Various network architectures, including CNNs \cite{bib18, bib17}, GANs \cite{bib31}, VAE \cite{bib24}, RNNs \cite{bib5, bib34}, and Transformers \cite{bib3, bib34} have been investigated. While auto-regressive models like RNNs and pure Transformers \cite{bib43} theoretically have the capacity to generate infinite motion, practical challenges such as mean regression arise. This phenomenon leads to motion ``freezing" or drifting into unnatural movements after several iterations. To address this, some studies \cite{bib43, bib44} propose periodic usage of the network's output as input during the training process. Additionally, Phase Function Neural Networks and their variants have been introduced \cite{bib45, bib46} to tackle the mean regression issue by conditioning network weights on the phase. However, their scalability in representing diverse movements is limited.

\textbf{Music-Driven Dance Generation} In recent years, data-driven deep learning has become the dominant technique for dance generation. Joao \cite{bib52} used graph convolutional networks to learn from a variety of dance datasets and generate new dance sequences that are smooth and continuous. This deep learning approach significantly improves the quality and continuity of the generated dances. Holden \cite{bib10}, Qiu \cite{bib11} and Starke \cite{bib28} built on deep learning by de-augmenting long-term dependency modeling as a means of generating more coherent long dance sequences. Common approaches include integrating skeleton information and employing attention mechanisms. Li \cite{bib7} built on that previous work by proposing the Full Attention Cross-Modal Transformer model (FACT), which can generate non-freezing, high-quality 3D motion sequences conditioned on music by learning audio-motion correspondences sequences. 

\textbf{Quaternion Networks} In various domains of deep learning such as few-shot segmentation \cite{bib65}, human motion synthesis \cite{bib64}, and multi-sensor signal processing, Quaternion Neural Networks have made significant strides. Similar to the task discussed in this paper, Quaternion representations are employed in neural network architectures as a parameterization for rotations.Quaternion networks, exemplified by QuaterNet \cite{bib64}, utilize quaternions to represent joint rotations in both Recurrent Neural Networks (RNNs) and Convolutional Neural Networks (CNNs). This approach addresses the discontinuity issues associated with Euler angles, achieving outstanding performance in long-term prediction tasks. In the context of our work, focused on music-driven dance generation, we propose constructing a learning process for the correlation between music and dance. This is essential as it requires consideration of the non-linear characteristics of both motion and music. Therefore, our method involves exploring the relationship between audio and motion features using quaternions. By leveraging quaternions, we aim to enhance the correlation between audio and motion, facilitating the generation of high-quality dance sequences.

\section{Methods}

\subsection{Overview of QEAN}\label{subsec1}\indent

In this paper, we propose a Quaternion-Enhanced Attention Network (QEAN) for generating high-quality dances under musical conditions, as illustrated in Fig. \ref{fig:2}.
  
We are given random motion seeds Y of length 120 frames and audio features Z of length 240 frames, where Y can be denoted as  $Y=\{y_1,y_2,\ldots,y_t\}$ and Z can be denoted as  $Z=\{z_1,z_2,\ldots,z_{t^\prime}\}$. Our objective is to generate a sequence of future motion from t+1 to $t^\prime$, $Y^\prime=\{y_{t+1} , y_{t+2} \ldots y_{t'}\}$, where $t^\prime\gg t$. QEAN first utilizes the two input transformers, the motion transformer $f_{mot}$ and the audio transformer $f_{audio}$, to encode features and generate motion and audio embeddings, represented as $h^y_{1:T}$ and $h^z_{1:T^\prime}$, respectively. Next, these two embedded features are combined and subjected to a quaternion parameterization operation (see 3.2 for details). This operation maps the features to four dimensions and embeds the information into the self-attention in a rotational manner using Spin Position Embedding (see 3.3 for details). Finally, the fused features are learned by a Quaternion Rotary Attention Transformer (see 3.4 for details) to generate the corresponding dance movements.

\subsection{Quaternion Algorithms and Quaternion Parameterization}\label{subsec2}\indent

We begin by elucidating the fundamental concepts of quaternions crucial for understanding the context of this paper. Quaternions, classified as hyper-complex numbers of rank 4, stand out as a direct and non-commutative extension of complex-valued numbers. In our proposed methodologies, the intricate interplay between Hamilton products and quaternion algebra emerges as the linchpin, forming the cornerstone of our innovative approaches. This exploration of  quaternion principles lays the groundwork for the subsequent discussions and applications detailed in this study.

A quaternion Q in quaternion domain D, Q $\in$ D, can be represented as:
    \begin{equation}
	  Q=e+f\textbf{i}+g\textbf{j}+h\textbf{k}
\end{equation}
Where e,f,g and h are real numbers,and \textbf{i},\textbf{j} and \textbf{k} are the quaternion unit basis.In a quaternion, e is the real part, where f\textbf{i}+g\textbf{j}+h\textbf{k}, with \(\textbf{i}^2\)=\(\textbf{j}^2\)=\(\textbf{k}^2\)=\textbf{ijk}=-1 is the imaginary part.

A pure quaternion is a quaternion whose real part is 0, resulting in the vector Q=f\textbf{i}+g\textbf{j}+h\textbf{k}. Operations on quaternions are defined as follows.

 The addition of two Quaternions is defined as:
 \begin{equation}
     Q+R=Q_e+R_e+(Q_f+R_f)\mathbf{i}+(Q_g+R_g)\mathbf{j}+(Q_h+R_h)\mathbf{k}
 \end{equation}
Where Q and P with subscripts denote the real and imaginary parts of the quaternions Q and P.

The Multiplication with scalar \(\gamma\) can be defined as:
  \begin{equation}
      \gamma Q=\gamma e+\gamma f\mathbf{i}+\gamma g\mathbf{j}+\gamma h\mathbf{k}
  \end{equation}

The conjugate complex $Q^\star$ of $Q$ can be defined as:
\begin{equation}
    Q^\star=e-f\mathbf{i}-g\mathbf{j}-h\mathbf{k}
\end{equation}

The multiplication of quaternions Q and R can be defined as follows: 
\begin{equation}
\begin{split}
    Q \bigotimes R = & (Q_eR_e - Q_fR_f - Q_gR_g - Q_hR_h) \\
    & + (Q_fR_e + Q_eR_f - Q_hR_g + Q_gR_h)\textbf{i} \\
    & + (Q_gR_e + Q_hR_e + Q_eR_g - Q_fR_h)\textbf{j} \\
    & + (Q_hR_e - Q_gR_f + Q_fR_g + Q_eR_h)\textbf{k}
\end{split}
\end{equation}

The  equation above clearly describes the exchange between quaternions Q and R, indicating that Hamiltonian product is essential in quaternion neural networks. In this study, we extensively employ the Hamiltonian product to learn the correlations between music and dance, which forms the foundation of QEAN and enhances its generalization ability.

To implement our approach, we combine music and motion features to create a feature vector. Specifically, 35-dimensional music features and 219-dimensional motion features can be combined into a 254-dimensional feature vector through concatenation, based on dimension and time correspondence.
Subsequently, we convert this concatenated feature vector into a sequence of quaternions. In this process, each music feature and three-dimensional dance motion feature are broken down into four components, representing a quaternion with a real part and three imaginary parts. As a result, the original 254-dimensional feature vector is transformed into a quaternion sequence with a length of 63 (disregarding the last two dimensions as they are insufficient to form a complete quaternion).
Finally, we input this quaternion sequence into the Spin Position Embedding for further processing. In this model, a position encoding is assigned to each quaternion, enabling the capture of position information within the sequence. 

By incorporating position information, the model gains a better understanding of the sequence and improves its performance accordingly.

\begin{figure}[htbp]
  \centering
  \includegraphics[width=1 \linewidth]{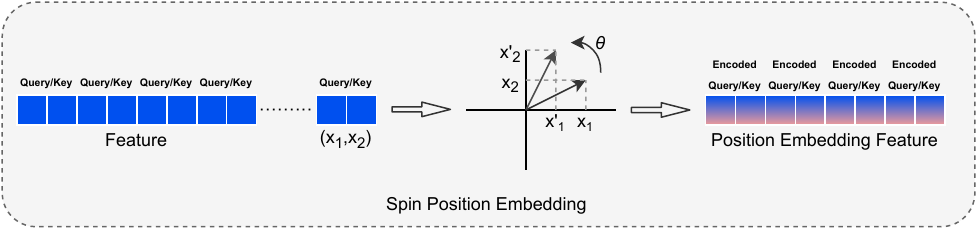}
  \caption{ The general situation of Spin Position Embedding. Specifically, the input action sequences and audio sequences in this paper are given feature vector representations after being encoded by their respective Transformers. The feature vectors of the action sequences are \(x_m\),and the feature vectors of the audio sequences are \(x_n\).These word vectors are then multiplied by different rotation matrices \(R_m\), \(R_n\) according to their positions m and n in their respective sequences to achieve the positional information of fusion. Finally, the encoded vectors of the action sequences are transformed into query vectors \(q_m\),and the rotationally transformed key vectors \(k_n\) of the encoded audio sequences are run on a click to compute the correlation between the two modal sequences. With this Spin Position Embedding, the modality can better model the positional information of the two sequences, as well as the correlation between them, thus increasing the learning of cross-modal representations.}
  \label{fig:3}
\end{figure}

\subsection{Spin Position Embedding}\label{subsec3}\indent

The main types of position embedding methods are relative position embedding and absolute position embedding methods. In 2017, the Transformer \cite{bib22} model was proposed. The concept of positional embedding was introduced in this model to provide information about the position of each word or token in this input sequence. This is crucial for NLP tasks that heavily rely on the relative position of words. In the Transformer model, positional embedding is used to encode information about different positions using sine and cosine functions, and the positional embedding is updated at different frequencies for different dimensions. In this way, the model is able to learn the relative positions of the tokens in the sequence. This way of position coding with sine and cosine, which is also known as absolute position coding, is easy to implement and relies directly on the absolute position without position loss, but this type of coding has poor generalisation ability, the model only adapts to a specific length of absolute position coding, and is prone to overfitting when the length varies, and performs poorly on tasks in long sequences. Relative positional embedding, on the other hand, is a method of obtaining positional embedding by using the relative distance or order relationship between lexical elements to rely on. This embedding method can provide relative position information between lexical elements instead of relying completely on absolute position. Such an embedding approach highlights the relevance of lexical elements in terms of content, which is conducive to content comprehension, and avoids the excessive  computation caused by the exponential growth of absolute positional embedding with position. Consequently, it improves the generalisation ability.
  
Motivated by the work of Jianlin Su \cite{bib9}, who proposed the  Rotary Position Embedding, a positional embedding method designed to enhance the performance of the Transformer architecture by integrating relative positional information into self-attention. The popular LLama2 \cite{bib58} model currently employs this position embedding approach.Therefore, we borrowed from Su and embedded the extracted audio and motion features into self-attention in the form of rotated positions to better learn the features in it and improve the computational efficiency. The basic idea can be seen in  Fig.\ref{fig:3} .

First, we define  a sequence of features of length N (since motion and audio features operate similarly in the process of SPE,The O in the next equation represents different operations for different eigenvectors in different situations): \(F_N=\{W_i\}_{i=1}^{N}\). Where \(w_i\) represents the i-th token in the input sequence, and the embedding corresponding to the input sequence \(F_N\) is denoted as \(E_N=\{x_i\}_{i=1}^{N}\), where \(x_i\) represents the d-dimensional embedding vector of the i-th token \(w_i\).

Before performing self-attention operations, we use the feature embedding vectors to calculate the q, k, and v vectors and incorporate the corresponding positional information. The function expressions are as follows:

\begin{align} q_s &= O_q(x_s,s) \\ k_t &= O_k(x_t,t) \\ v_t &= O_v(x_t,t)\end{align}

Here, \(q_s\) represents the query vector for the s-th token with positional information s integrated into the feature vector \(x_s\), while \(k_t\) and \(v_t\) represent the key and value vectors for the t-th token with positional information t integrated into the feature vector \(x_t\).

Then, we need to compute the output of self-attention for the s-th feature embedding vector \(x_s\) . This involves calculating an attention score between \(q_s\) and other \(k_t\) , and then multiplying the attention score by the corresponding \(v_t\) , followed by summation to obtain the output vector \(o_s\):

\begin{align}      
  a_{s,t}=\frac{exp(\frac{q_s^{T}k_t}{\sqrt{d}})}{\sum_{j=1}^{N} exp(\frac{q_sk_j}{\sqrt{d}})}\\
  o_s=\sum_{n=1}^{N} a_{s,t}v_n
\end{align}

Next, in order to leverage the relative positional relationships between the mentioned tokens, let's assume that the dot product operation between the query vector \(q_s\) and the key vector \(k_t\) is represented by a function g . The input to function g includes the word embedding vectors \(x_s\), \(x_t\) and their relative position s-t:

\begin{align}     
  <O_q(x_s,s),O_k(x_t,t)>=g(x_s,x_t,s-t)
\end{align}  
    
We then discover an alternative approach to position embedding that upholds the aforementioned relationship.

\begin{align}   
  &O_q(x_s,s)=(W_qx_s)e^{is\theta}  \\
  &O_k(x_t,t)=(W_kx_t)e^{it\theta} \\
  &O(x_s,x_t,s-t)= Re[(W_qx_s)(W_kx_t)^{*}e^{i(s-t)\theta}]
\end{align} 

Here, \textbf{x} represents any real number, \textbf{e} is the base of the natural logarithm, and \textbf{i} is the imaginary unit in complex numbers.

We can cleverly use Euler's formula \(e^{ix}=cosx+isinx\), where the real part is cosx and the imaginary part sinx is of a complex number.
  
After transformation, formulas \textbf{O} and  \textbf{g} can be changed to:

\begin{align}     
  &e^{is\theta}=cos(s\theta)+isin(s\theta)\\      
  &e^{it\theta}=cos(t\theta)+isin(t\theta)      \\
  &e^{i(s-t)\theta}=cos((s-t)\theta)+isin((s-t)\theta)     \\ 
  &O_q(x_s,s)=(W_qx_s)e^{is\theta}
\end{align} 

  Then, according to linear algebra, we can represent \(q_s\) using a matrix:
\begin{align}       
  &q_s=\begin{pmatrix} q_s^{(1)} \\ q_s^{(2)} \end{pmatrix}=(W_qx_s)=\begin{pmatrix} W_q^{(11)} & W_q^{(12)} \\ W_q^{(21)} & W_q^{(22)}\end{pmatrix}\begin{pmatrix} x_s^{(1)} \\ x_s^{(2)} \end{pmatrix}\\   
  &O_q(x_s,s)=(W_qx_s)e^{is\theta}=q_se^{is\theta}
\end{align}

Therefore, multiplying these two complex numbers, we get the following result:

\begin{align}
q_se^{is\theta}=[q_s^{(1)}cos(s\theta)-q_s^{(2)}sin(s\theta) ,q_s^{(2)}cos(s\theta)+q_s^{1}sin(s\theta)
\end{align} 
     
Then, we magically discover that the above expression is equal to the query vector multiplied by a rotation matrix:
  
\begin{align}
  O_q(x_s,s) & = (W_qx_s)e^{is\theta} = q_se^{is\theta} \nonumber \\
            & = \begin{pmatrix} \cos(s\theta) & -\sin(s\theta) \\ \sin(s\theta) & \cos(s\theta)\end{pmatrix}\begin{pmatrix} q_s^{(1)} \\ q_s^{(2)}\end{pmatrix}
\end{align}

Similarly, the key vector \(k_t\) can be represented as follows:

\begin{align}
  O_k(x_t,t) &= (W_kx_t)e^{in\theta} = k_te^{it\theta} \nonumber \\
             & = \begin{pmatrix} \cos(t\theta) & -\sin(t\theta) \end{pmatrix} \begin{pmatrix} k_t^{(1)} \\ k_t^{(2)} \end{pmatrix} \nonumber \\
             &\quad + \begin{pmatrix} \sin(t\theta) & \cos(t\theta)\end{pmatrix}\begin{pmatrix} k_t^{2} \\ k_t^{1} \end{pmatrix}
\end{align}

By rearranging the above formulas, we can simplify the following expression:

\begin{align}
\small
&<O_q(x_s,s),O_k(x_t,t)> \nonumber \\
&= \left(\begin{pmatrix} \cos(s\theta) & -\sin(s\theta) \\ \sin(s\theta) & \cos(s\theta)\end{pmatrix}^T \begin{pmatrix} q_s^{(1)} \\ q_s^{(2)} \end{pmatrix}\right)^T \nonumber \\
&\quad \begin{pmatrix} \cos(t\theta) & -\sin(t\theta) \\ \sin(t\theta) & \cos(t\theta)\end{pmatrix}\begin{pmatrix} k_t^{(1)} \\ k_t^{(2)} \end{pmatrix} \nonumber \\
&= \begin{pmatrix} q_s^{(1)} & q_s^{(2)} \end{pmatrix}\begin{pmatrix} \cos((s-t)\theta) & -\sin((s-t)\theta) \\ \sin((s-t)\theta) & \cos((s-t)\theta)\end{pmatrix}\begin{pmatrix} k_t^{(1)} \\ k_t^{(2)} \end{pmatrix}
\end{align}
  
With the above formulas, we can summarize the following calculation process: In simple terms, the process of self-attention with Spin Position Embedding involves, for each feature embedding vector in the token sequence, first calculating its corresponding query and key vectors. Then, for each token position, calculate the corresponding rotated position embedding information. After that, apply the rotation transformation to the elements of the query and key vectors for each token position in pairs, and finally, calculate the dot product between the query and key to obtain the result of self-attention.

\subsection{Quaternion Rotary Attention}\label{subsec4}

\begin{figure}[htbp]
  \centering
  \includegraphics[width=1 \linewidth]{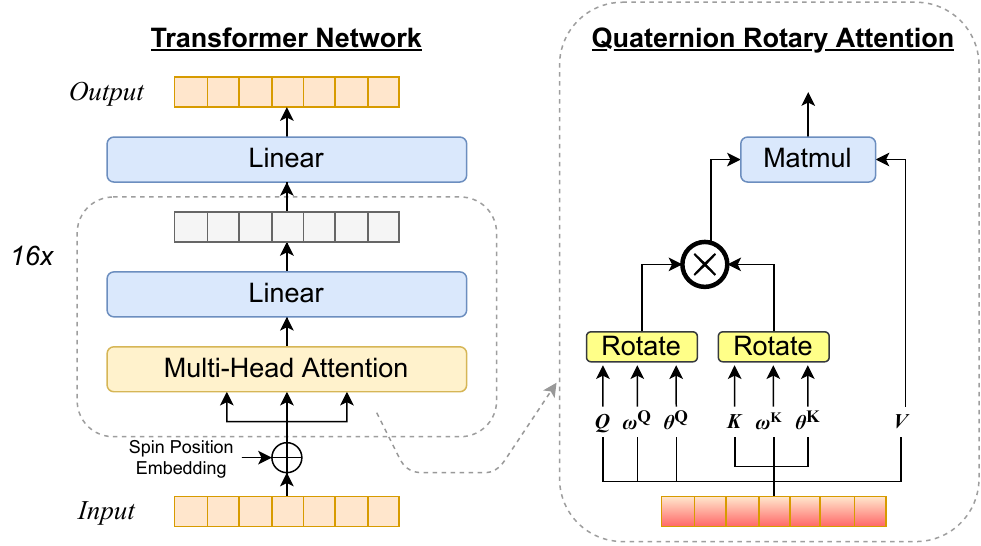}
  \caption{The overall of our Transformer structure.Our Transformer structure enhances the generalisation ability of the model by adding regularisation means such as Dropout in multiple places and adjusting the number of Attention heads to expand the model capacity on the basis of the original. The absolute position information of the input sequence is converted into a polar coordinate representation of the relative position using Spin Position Embedding, ($\rho$, $\theta$) where $\rho$ denotes the distance from the centre point, $\theta$ denotes the relative angle. This Spin Position Embedding module provides better local relative positions with some rotational invariance. In this way, our model can better support some tasks that are sensitive to position information, such as behavioural sequence modelling and 3D shape analysis. }
  \label{fig:4}
  
\end{figure}

For the features after the rotated attention module, we assume that there are N-length query series $X$  and an M-length key-values series $\gamma$. Firstly the original $\chi$ and $\gamma$ are projected onto the  representation space,and  a series of operations are performed: $Q=\chi{W^Q}\in R^{N\times d}$, $K=\gamma{W^K}\in R^{M\times d}$ and $V=\gamma{W^V}\in R^{M\times d}$. 

Here, Q represents the query vector, K represents the key, V represents the value, d represents the number of channels in the attention layer and W represents the trainable weights. Then, QRA will calculate $H=Attn(X,\gamma)$ to the map query series to output $H$ using key-value series.

Frequency/phase-Generation:
\begin{align}
    \begin{gathered}\left.\left(\begin{array}{c}\omega_{1}^\mathrm{Q}\\\cdots\\\omega_{P}^\mathrm{Q}\end{array}\right.\right),\left(\begin{array}{c}\theta_{1}^\mathrm{Q}\\\cdots\\\theta_{P}^\mathrm{Q}\end{array}\right) =\mathrm{Conv}(Q;W_{\omega}^{\mathrm{Q}}),\mathrm{Conv}(Q;W_{\theta}^{\mathrm{Q}}), \\\left.\left(\begin{array}{c}\omega_{1}^\mathrm{K}\\\cdots\\\omega_{P}^\mathrm{K}\end{array}\right.\right),\left(\begin{array}{c}\theta_{1}^\mathrm{K}\\\cdots\\\theta_{P}^\mathrm{K}\end{array}\right) =\mathrm{Conv}(K;W_{\omega}^{\mathrm{K}}),\mathrm{Conv}(K;W_{\theta}^{\mathrm{K}}), \end{gathered}
\end{align}

Series-Rotation 
\begin{align}
    \begin{aligned}\Phi_p(Q,\text{pos}^\mathrm{Q})&=\tilde{Q}e^{\mathbf{i}(2\pi\omega_p^\mathrm{Q}\text{pos}^\mathrm{Q}+\theta_p^\mathrm{Q})},\quad p=1,2,\cdots,P\\\Psi_p(K,\text{pos}^\mathrm{K})&=\tilde{K}e^{\mathbf{j}(2\pi\omega_p^\mathrm{K}\text{pos}^\mathrm{K}+\theta_p^\mathrm{K})},\quad p=1,2,\cdots,P\end{aligned}
\end{align}

Series-Attention with softmax-kernel (shown in Fig.5)
\begin{align}
    S=\text{softmax}\left(\frac{1}{P\sqrt{d}}\sum_{p=1}^{P}\text{Re}[\Phi_{p}(Q,\text{pos}^{\mathbb{Q}})\Psi_{p}(K,\text{pos}^{\mathbb{K}})^{\mathsf{H}}]\right)
\end{align}

Series-Aggregation:
\begin{align}
    H=SV
\end{align}

\begin{figure*}[htbp]
  \centering
  \includegraphics[width=0.35\linewidth]{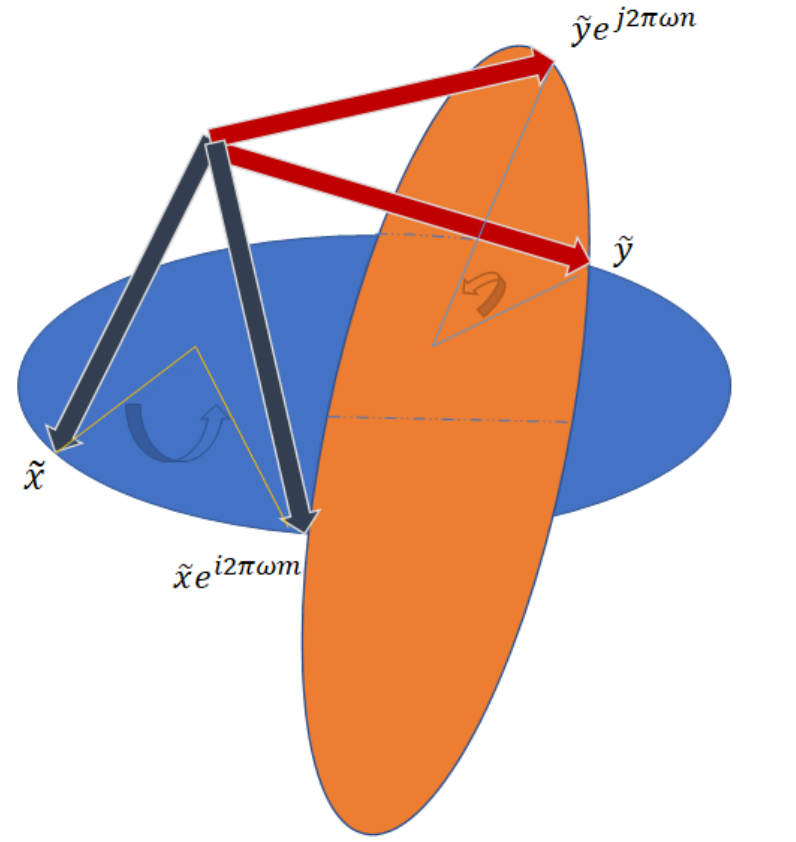} 
  \caption{A three-dimensional illustration of a rotated softmax-kernel. The rotated softmax-kernel represents the embeddings in quaternion form and rotates them using the angular frequency \(\omega\) .Thus, embeddings with different phases can be distinguished. Finally, the similarity of the rotated embeddings is measured by measuring the exponential dot product between them.}
  \label{fig:5}
\end{figure*}

Here, we hypothesize that the series have $P$ periods, and $P$ is a hyper-parameter . In frequency/phase-generation step, we utilize 1D convolutions with activation ReLU to generate $P$ latent frequencies $\omega_{1\sim P}^{\mathrm{Q}}\in[0,+\infty)^{N\times1}$($\omega_{1{\sim}P}^{\mathrm{K}}$ is similar). Convolutions can effectively capture local contexts of each time step to generate reliable latent frequencies, and these latent frequencies are not identical at each time step implying variable periods. Moreover, to account for phase shifts, we additionally generate $P$ latent phases  $\theta_{1\sim P}^{\mathrm{Q}}\in(-\pi,\pi)^{N\times1}$ using 1D convolutions with activation $\pi\cdot\mathrm{tanh}$
($\theta_{1{\sim}P}^{\mathrm{K}}$ is similar).In series-rotation step, we rotate the representations at each time step according to the learned latent frequencies and phases in the previous step.Each row vector of $\tilde{Q},\tilde{K}$ is in the quaternion form of the corresponding row vector of Q, k, and $pos^{Q}=[0,1,2,\cdots,N-1]^{\mathrm{T}}/\mathrm{N}$,$pos^{K}=[0,1,2,\cdots,M-1]^{\mathrm{T}}/\mathrm{M}$ are position vectors of series Q and K, respectively. In the series-attention step, to integratedly capture position-wise similarity under multiple periods, the unnormalized similarity is the mean of quaternion dot-product under multiple rotations. Finally, in the series-aggregation step, the outputs $H\in\mathbb{R}^{N\times d}$ is generated using the softmax-normalized similarity. In practice, we employ the multi-head variant of QRA, and will not go into details here, as it can be derived quite directly. Notice that, QRA is more expressive than canonical dot-product attention. When $P$= 1, $\omega$ = 0 and $\theta$ = 0, QRA degenerates into canonical attention.

\section{Experiments}

\subsection{Datasets}\label{subsec5}

  The AIST++ \cite{bib8} dance movement dataset was constructed from the AIST dance \cite{bib7} video database. A well-developed process was designed for estimating camera parameters, 3D human keypoints and 3D human dance movement sequences from multi-view videos. The dataset provides 3D human keypoint annotations and camera parameters for 10.1 million images covering 30 different subjects in 9 viewpoints. These features make it the largest and richest dataset containing 3D human keypoint annotations currently available. Additionaly, the dataset contains 1,408 3D human dance movement sequences represented as joint rotations and root trajectories. These dance movements are evenly distributed across 10 dance genres and contain hundreds of choreographies. The duration of the movements ranges from 7.4 to 48.0 seconds, and each dance movement is accompanied by corresponding music. Based on these annotations, AIST++ is designed to support multiple tasks including multi-view human keypoint estimation, human motion prediction/generation, and cross-modal analysis between human motion and music.

\subsection{Implementation Details}\label{subsec6}

 In our primary experiments, the model takes a seed motion sequence spanning 120 frames (2 seconds) and a music sequence covering 240 frames (4 seconds) as input. These two sequences are aligned at the initial frame, and the model's output consists of a future motion sequence with N=20 frames supervised by L2 loss. During the inference process, future motions are continuously generated in an auto-regressive manner at 60 FPS, with only the first predicted motion retained at each step.For music feature extraction, we employ the publicly available audio processing toolbox, Librosa \cite{bib47}, which includes 1-dimensional envelope, 20-dimensional MFCC, 12-dimensional chroma, 1-dimensional one-hot peaks, and 1-dimensional one-hot beats, resulting in a 35-dimensional music feature. The motion features combine a 9-dimensional representation of rotation matrices for all 24 joints with a 3-dimensional global translation vector, resulting in a 219-dimensional motion feature. These raw audio and motion features are initially embedded into 800-dimensional hidden representations using linear layers, with learnable position embeddings added before inputting them into the transformer layers. All three transformers (audio, motion, cross-modal) feature 16 attention heads with a hidden size of 800. In terms of training details, all experiments are trained using the Adam optimizer with a batch size of 16. The learning rate starts at 1e-4 and decays to \{1e-5,1e-6\} at \{90k, 150k\} steps, respectively. Training concludes after 500k steps, taking approximately 2 days on one RTX 3090. The baseline comparison includes the latest works on 3D dance generation with music and seed motion as input, such as  Li \cite{bib7} and Li et al \cite{bib3}. For a more comprehensive evaluation, we also compare it with the recent state-of-the-art 2D dance generation method, DanceRevolution \cite{bib4}. We adapt this work to output 3D joint positions for a direct quantitative comparison with our results, even though joint positions do not allow for immediate repositioning. The official code provided by the authors is used to train and test these baselines on the same dataset as ours.

\subsection{Quanitative Evalutation}

In this section, we assess the performance of our proposed Multi-modal Roformer across three key dimensions: (1) motion quality (2) generation diversity and (3) motion-music correlation. The results presented in Table 1 demonstrate that, under identical experimental conditions, our model surpasses state-of-the-art methods \cite{bib2, bib5, bib6} in these aspects.

\textbf{Motion Quality}: 
Similar to previous studies, we assess the quality of generated motion by computing the Frechet Inception distance (FID) \cite{bib49}, which measures the dissimilarity between the distribution of generated motion and ground-truth motion. To capture motion features, we utilize two meticulously crafted motion feature extractors, as undisclosed motion encoders were employed in earlier works \cite{bib36}. These extractors include: (1) a geometric feature extractor, generating a boolean vector that represents geometric relationships among specific body points in the motion sequence, and (2) a dynamic feature extractor, mapping the motion sequence to capture dynamic aspects such as velocity and acceleration.We designate FID based on these geometric and dynamic features as \(FID_g\) and \(FID_d\), respectively. The metrics are computed by comparing real dance motion sequences in the AIST++ test set with 40 generated motion sequences, each comprising T = 1200 frames (20 seconds). As depicted in Table 1, our generated motion sequences exhibit distributions that are closer to ground-truth motion compared to the three methods.

\textbf{Generation Diversity}:We also assess the model's capacity to generate diverse dance movements in response to different input music, comparing its performance to baseline methods. Following a methodology similar to previous research \cite{bib37}, we compute the average Euclidean distance in the feature space of 40 generated motions from the AIST++ test set to quantify diversity. The motion diversity in geometric and dynamic feature spaces is denoted as \(Dist_g\) and \(Dist_k\), respectively.Table1 illustrates that our method excels in generating more diverse dance movements in comparison to the baselines, with the exception of Li \cite{bib31}. The latter discretizes motions, resulting in discontinuous outputs and elevated \(Dist_k\).

\textbf{Motion-Music Correlation}:Moreover, we gauge the correlation between the generated 3D motion and input music by introducing a novel metric known as the Beat Alignment Score. This metric evaluates the motion-music correlation by measuring the similarity between the beats in the motion and music. Librosa \cite{bib47} is employed to extract music beats, while motion beats are computed as local minima in the motion velocity. The Beat Alignment Score is articulated as the average distance between each motion beat and its nearest music beat. To be specific, our Beat Alignment Score is defined as:

\begin{align}
BeatAlign=\frac{1}{z}\sum_{i=1}^{z} exp(-\frac{min\forall{t_j^d\in B^d||t_i^c-t_j^d||^2}}{2\alpha^2})
\end{align}
    where $B^c=\left\{{t_i^{c}}\right\}$
 is the set of motion beats, $B^d=t_j^{d}$ is the music beats, and $\alpha$ is a parameter for normalizing sequences with different FPS.
    
    We set $\alpha=3$ for all experiments since the FPS for all our experimental sequences is 60. A similar metric called Beat Hit Rate is introduced in, but it relies on manually set thresholds for alignment (``hits") depending on the dataset, while our metric directly measures distances. This metric is explicitly designed to be unidirectional, as dance movements do not necessarily need to match every music beat. On the other hand, each dynamic beat should have a corresponding music beat. To calibrate the results, we compute correlation metrics for the entire AIST++ dataset (upper bound) and randomly paired data (lower bound). As shown in Table \ref{table1}, our generated motion shows better correlation with input music compared to the baselines. However, there is still considerable room for improvement for all methods, including ours, when compared to actual data. This reflects that music-motion correlation remains a challenging problem.

\begin{table}[htbp]
  \centering
    \begin{tabular}{c|cc|cc|c}
    \toprule
    \multirow{2}[4]{*}{\textbf{Methods}} & \multicolumn{2}{c|}{\textbf{Motion Quality}} & \multicolumn{2}{c|}{\textbf{Motion Diversity}} & \multicolumn{1}{c}{\textbf{Motion-Music Corr}} \\
\cmidrule{2-6}    \multicolumn{1}{c|}{} & \textbf{FID\textsubscript{k}↓} & \textbf{FID\textsubscript{g}↓} & \textbf{Dist\textsubscript{k}↑} & \textbf{Dist\textsubscript{g}↑} & \multicolumn{1}{c}{\textbf{BeatAlign}↑} \\
    \midrule
    AIST++ & - & - & 9.057 & 7.556 & 0.292 \\ \hline
            AIST++(random) & - & - & - & - & 0.213 \\ \hline
            Li et al\cite{bib4}. & 86.43 & 20.58 & 6.85 & 4.93 & 0.232 \\ \hline
            Dancenet\cite{bib5} & 69.18 & 17.76 & 2.86 & 2.72 & 0.232 \\ \hline
            DanceRevolution\cite{bib6} & 73.42 & 31.01 & 3.52 & 2.46 & 0.22 \\ \hline
            FACT(baseline)\cite{bib7} & 48.95 & 28.1 & 4.9 & 6.69 & 0.232 \\ \hline
           \bfseries our  & \bfseries 30.1 & \bfseries 11.5 & \bfseries 7.82 & \bfseries 9.37 & \bfseries 0.239 \\ \hline
    \bottomrule
    \end{tabular}%
    \caption{\textbf{ Conditional Motion Generation Evaluation on AIST++ dataset}. Comparing to the three recent state-of-art methods, our generated motions are more realistic, better correlated with input music and more diversified.}
  \label{table1}%
\end{table}%

\subsection{Ablation Study}
  We conducted ablation studies on the Spin Position Embedding and Multi-modal Quaternion parameterization, respectively. The quantitative scores are shown in \ref{table2}.

\textbf{Position Embedding}   
In the ablation experiments focused on position coding, we explore two distinct approaches and conduct experiments based on the following configurations: (1) a learnable coding approach for absolute positions (baseline), and (2) a rotary coding approach for relative positions. Method 2 was selected to introduce explicit relative position dependence in the self-attention formulation. This choice offers increased flexibility in sequence length, a potential reduction in dependencies between tokens, and the capacity to encode relative positions for linear self-attention.As illustrated in Table \ref{table2}, we observe that the rotational embedding method of relative position leads to a significant reduction in the \(FID_g\) values compared to the original method. This indicates that dances generated using rotary position embedding are notably closer to reality.

\textbf{Quaternion parameterization}  Here, we performed ablation experiments on the original baseline as well as with the addition of the QRA module. Through Table \ref{table2}, we can observe that Quaternion Rotary Attention (QRA), by introducing quaternion operations, is able to fully explore the relationship between audio and motion, and achieves more significant enhancement results.

\begin{table}[!ht]
    \centering
    \resizebox{1\textwidth}{!}{
    \begin{tabular}{|l|l|l|l|l|l|}
    \hline       
        ~ & \(FID_{k}\) ↓ & \(FID_{g}\) ↓  & BeatAlign ↑ \\ \hline    
        baseline & 48.95 & 28.1  & 0.232 \\ \hline
        \bfseries baseline+Spin Position Embedding  & \bfseries 30.1 & \bfseries 11.5 & \bfseries 0.239 \\ \hline            
        \bfseries baseline+QRA  & \bfseries 46.33 & \bfseries 26.2 & \bfseries 0.236 \\ \hline
    \end{tabular}
    }
    \begin{flushleft}
    \caption{\textbf{Ablation Study on Spin Position Embedding and Quaternion Rotary Attention}.As illustrated in the table, the experimental results clearly demonstrate the effectiveness of our proposed method. The graph shows a significant improvement in performance metrics when compared to the baseline approach.}
    \label{table2}
    \end{flushleft}
\end{table}

\begin{figure*}[htbp]
  \centering
  \includegraphics[width=1.0 \linewidth]{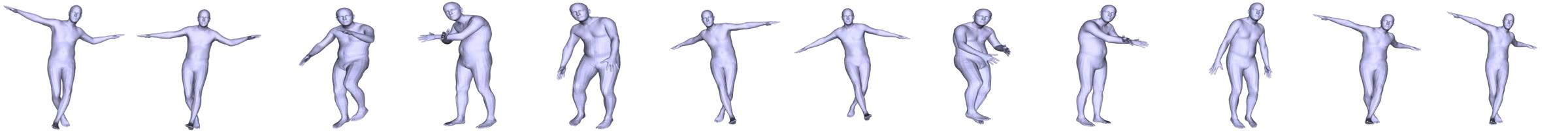} 
  \caption{Frame extraction. The visual representation clearly emphasizes the effectiveness of our proposed method.}
  \label{fig:6}
\end{figure*}

\begin{figure*}[htbp]
  \centering
  \includegraphics[width=1.0 \linewidth]{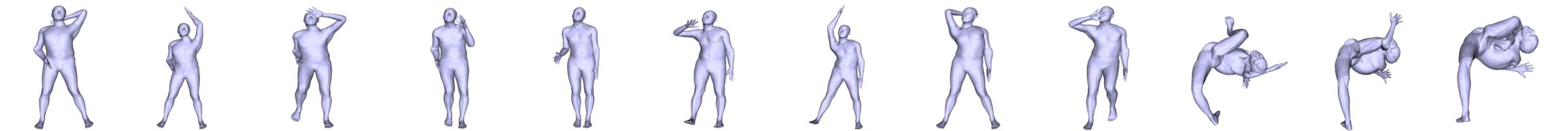}
  \caption{This image illustrates the frame extraction effect of a dance generated by alternative methods. During the post-production phase, the generated dance movements exhibit phenomena of dance collapse and unscientific limb folding.}
  \label{fig:7}
\end{figure*}
\newpage

\section{Conclusion}\label{sec5}\indent

We propose a network called QEAN for generating 3D dance movements. QEAN employs Spin Position Embedding (SPE) the position encoding part to embed the position information in a rotational manner in the self-attention, which improves the model's representation of the sequences and enhances the model's understanding of the human movements in terms of their temporal order. Additionally, we propose Quaternion Rotary Attention (QRA), a quaternion-valued relational learning network, which uses quaternion values to explore the temporal coordination between music and dance. To demonstrate the superiority of QEAN, we conducted experiments on the AIST++ dataset. The results of the relevant experimental data demonstrate the superiority of our approach in the 3D dance generation task. Furthermore, the results of our ablation experiments illustrate the importance of SPE and QRA in this task.
  
\section{Acknowledgement}\label{sec6}
This work was supported in part by National Natural Science Foundation under Grant 92267107, the Science and Technology Planning Project of Guangdong under Grant 2021B0101220006,  Science and Technology Projects in Guangzhou under Grant 202201011706, Key Areas Research and Development Program of Guangzhou under Grant 2023B01J0029, Science and technology research in key areas in Foshan under Grant 2020001006832, Key Area Research and Development Program of Guangdong Province under Grant 2018B010109007 and 2019B010153002, Science and technology projects of Guangzhou under Grant 202007040006, and Guangdong Provin-cial Key Laboratory of Cyber-Physical System under Grant 2020B1212060069.

\section{Declarations}
\textbf{Conflict of interest} We declare that we do not have any commercial or associative interest that represents a conflict of interest in connection
with the work submitted.

\nocite{*}

\printbibliography

@article{bib1,
author = {Yang, Yue and Zhang, Ensi},
year = {2023},
month = {12},
pages = {315-328},
title = {Cultural thought and philosophical elements of singing and dancing in Indian films},
volume = {46},
journal = {Trans/Form/Ação},
doi = {10.1590/0101-3173.2023.v46n4.p315}
}

@article{bib2,
  title={A citation analysis of business librarianship: Examining the Journal of Business and Finance Librarianship from 1990–2014},
  author={Mark Siciliano},
  journal={Journal of Business \& Finance Librarianship},
  year={2017},
  volume={22},
  pages={81 - 96},
  url={https://api.semanticscholar.org/CorpusID:63474056}
}

@article{bib3,
  title={Learning to Generate Diverse Dance Motions with Transformer},
  author={Jiaman Li and Yihang Yin and Hang Chu and Yi Zhou and Tingwu Wang and Sanja Fidler and Hao Li},
  journal={ArXiv},
  year={2020},
  volume={abs/2008.08171},
  url={https://api.semanticscholar.org/CorpusID:221173065}
}

@inproceedings{bib4,
  title={Dance Revolution: Long-Term Dance Generation with Music via Curriculum Learning},
  author={Ruozi Huang and Huang Hu and Wei Wu and Kei Sawada and Mi Zhang and Daxin Jiang},
  booktitle={International Conference on Learning Representations},
  year={2020},
  url={https://api.semanticscholar.org/CorpusID:235614403}
}

@article{bib5,
  title={Dance Generation with Style Embedding: Learning and Transferring Latent Representations of Dance Styles},
  author={Xinjian Zhang and Yi Xu and Su Yang and Longwen Gao and Huyang Sun},
  journal={ArXiv},
  year={2021},
  volume={abs/2104.14802},
  url={https://api.semanticscholar.org/CorpusID:233476346}
}

@inproceedings{bib6,
  title={Dance Revolution: Long-Term Dance Generation with Music via Curriculum Learning},
  author={Ruozi Huang and Huang Hu and Wei Wu and Kei Sawada and Mi Zhang and Daxin Jiang},
  booktitle={International Conference on Learning Representations},
  year={2020},
  url={https://api.semanticscholar.org/CorpusID:235614403}
}

@article{bib7,
  title={AI Choreographer: Music Conditioned 3D Dance Generation with AIST++},
  author={Ruilong Li and Sha Yang and David A. Ross and Angjoo Kanazawa},
  journal={2021 IEEE/CVF International Conference on Computer Vision (ICCV)},
  year={2021},
  pages={13381-13392},
  url={https://api.semanticscholar.org/CorpusID:236882798}
}

@inproceedings{bib8,
  title={AIST Dance Video Database: Multi-Genre, Multi-Dancer, and Multi-Camera Database for Dance Information Processing},
  author={Shuhei Tsuchida and Satoru Fukayama and Masahiro Hamasaki and Masataka Goto},
  booktitle={International Society for Music Information Retrieval Conference},
  year={2019},
  url={https://api.semanticscholar.org/CorpusID:208334750}
}

@article{bib9,
  title={RoFormer: Enhanced Transformer with Rotary Position Embedding},
  author={Jianlin Su and Yu Lu and Shengfeng Pan and Bo Wen and Yunfeng Liu},
  journal={ArXiv},
  year={2021},
  volume={abs/2104.09864},
  url={https://api.semanticscholar.org/CorpusID:233307138}
}

@article{bib10,
  title={A deep learning framework for character motion synthesis and editing},
  author={Daniel Holden and Jun Saito and Taku Komura},
  journal={ACM Transactions on Graphics (TOG)},
  year={2016},
  volume={35},
  pages={1 - 11},
  url={https://api.semanticscholar.org/CorpusID:18149328}
}

@article{bib11,
  title={Cross View Fusion for 3D Human Pose Estimation},
  author={Haibo Qiu and Chunyu Wang and Jingdong Wang and Naiyan Wang and Wenjun Zeng},
  journal={2019 IEEE/CVF International Conference on Computer Vision (ICCV)},
  year={2019},
  pages={4341-4350},
  url={https://api.semanticscholar.org/CorpusID:201891326}
}

@article{bib12,
  title={PCMG:3D point cloud human motion generation based on self-attention and transformer},
  author={Weizhao Ma and Mengxiao Yin and Guiqing Li and Feng Yang and Kan Chang},
  journal={The Visual Computer},
  year={2023},
  url={https://api.semanticscholar.org/CorpusID:261566852}
}

@inproceedings{bib17,
  title={EasyPhoto: Your Smart AI Photo Generator},
  author={Ziheng Wu and Jiaqi Xu and Xinyi Zou and Kunzhe Huang and Xing Shi and Jun Huang},
  year={2023},
  url={https://api.semanticscholar.org/CorpusID:263829612}
}

@inproceedings{bib18,
  title={Predicting Head Pose from Speech with a Conditional Variational Autoencoder},
  author={David Greenwood and Stephen D. Laycock and Iain Matthews},
  booktitle={Interspeech},
  year={2017},
  url={https://api.semanticscholar.org/CorpusID:11113871}
}

@inproceedings{bib22,
  title={Attention is All you Need},
  author={Ashish Vaswani and Noam M. Shazeer and Niki Parmar and Jakob Uszkoreit and Llion Jones and Aidan N. Gomez and Lukasz Kaiser and Illia Polosukhin},
  booktitle={Neural Information Processing Systems},
  year={2017},
  url={https://api.semanticscholar.org/CorpusID:13756489}
}

@article{bib24,
  title={Cross-Conditioned Recurrent Networks for Long-Term Synthesis of Inter-Person Human Motion Interactions},
  author={Jogendra Nath Kundu and Himanshu Buckchash and Priyanka Mandikal and Rahul},
  journal={2020 IEEE Winter Conference on Applications of Computer Vision (WACV)},
  year={2020},
  pages={2713-2722},
  url={https://api.semanticscholar.org/CorpusID:214675800}
}

@article{bib25,
  title={Scheduled Sampling for Sequence Prediction with Recurrent Neural Networks},
  author={Samy Bengio and Oriol Vinyals and Navdeep Jaitly and Noam M. Shazeer},
  journal={ArXiv},
  year={2015},
  volume={abs/1506.03099},
  url={https://api.semanticscholar.org/CorpusID:1820089}
}

@article{bib28,
  title={Quantized GAN for Complex Music Generation from Dance Videos},
  author={Ye Zhu and Kyle Olszewski and Yuehua Wu and Panos Achlioptas and Menglei Chai and Yan Yan and S. Tulyakov},
  journal={ArXiv},
  year={2022},
  volume={abs/2204.00604},
  url={https://api.semanticscholar.org/CorpusID:247922422}
}

@article{bib31,
  title={Feel The Music: Automatically Generating A Dance For An Input Song},
  author={Purva Tendulkar and Abhishek Das and Aniruddha Kembhavi and Devi Parikh},
  journal={ArXiv},
  year={2020},
  volume={abs/2006.11905},
  url={https://api.semanticscholar.org/CorpusID:219572850}
}

@article{bib34,
  title={Bailando: 3D Dance Generation by Actor-Critic GPT with Choreographic Memory},
  author={Lian Siyao and Weijiang Yu and Tianpei Gu and Chunze Lin and Quan Wang and Chen Qian and Chen Change Loy and Ziwei Liu},
  journal={2022 IEEE/CVF Conference on Computer Vision and Pattern Recognition (CVPR)},
  year={2022},
  pages={11040-11049},
  url={https://api.semanticscholar.org/CorpusID:247627867}
}

@article{bib35,
  title={Genre-Conditioned Long-Term 3D Dance Generation Driven by Music},
  author={Yuhang Huang and Junjie Zhang and Shuyan Liu and Qian Bao and Dan Zeng and Zhineng Chen and Wu Liu},
  journal={ICASSP 2022 - 2022 IEEE International Conference on Acoustics, Speech and Signal Processing (ICASSP)},
  year={2022},
  pages={4858-4862},
  url={https://api.semanticscholar.org/CorpusID:249437513}
}

@inproceedings{bib36,
  title={FMDistance: A Fast and Effective Distance Function for Motion Capture Data},
  author={Kensuke Onuma and Christos Faloutsos and Jessica K. Hodgins},
  booktitle={Eurographics},
  year={2008},
  url={https://api.semanticscholar.org/CorpusID:8323054}
}

@inproceedings{bib37,
  title={LXMERT: Learning Cross-Modality Encoder Representations from Transformers},
  author={Hao Hao Tan and Mohit Bansal},
  booktitle={Conference on Empirical Methods in Natural Language Processing},
  year={2019},
  url={https://api.semanticscholar.org/CorpusID:201103729}
}

@article{bib39,
  title={Long Short-Term Memory},
  author={Sepp Hochreiter and J{\"u}rgen Schmidhuber},
  journal={Neural Computation},
  year={1997},
  volume={9},
  pages={1735-1780},
  url={https://api.semanticscholar.org/CorpusID:1915014}
}

@article{bib40,
  title={Convolutions Die Hard: Open-Vocabulary Segmentation with Single Frozen Convolutional CLIP},
  author={Qihang Yu and Ju He and Xueqing Deng and Xiaohui Shen and Liang-Chieh Chen},
  journal={ArXiv},
  year={2023},
  volume={abs/2308.02487},
  url={https://api.semanticscholar.org/CorpusID:260611350}
}

@article{bib42,
  title={Multimodal Transformer for Unaligned Multimodal Language Sequences},
  author={Yao-Hung Hubert Tsai and Shaojie Bai and Paul Pu Liang and J. Zico Kolter and Louis-Philippe Morency and Ruslan Salakhutdinov},
  journal={Proceedings of the conference. Association for Computational Linguistics. Meeting},
  year={2019},
  volume={2019},
  pages={6558-6569},
  url={https://api.semanticscholar.org/CorpusID:173990158}
}

@article{bib43,
  title={VALUE: A Multi-Task Benchmark for Video-and-Language Understanding Evaluation},
  author={Linjie Li and Jie Lei and Zhe Gan and Licheng Yu and Yen-Chun Chen and Rohith Krishnan Pillai and Yu Cheng and Luowei Zhou and Xin Eric Wang and William Yang Wang and Tamara L. Berg and Mohit Bansal and Jingjing Liu and Lijuan Wang and Zicheng Liu},
  journal={ArXiv},
  year={2021},
  volume={abs/2106.04632},
  url={https://api.semanticscholar.org/CorpusID:235377363}
}

@article{bib44,
  title={Learning Human Motion Models for Long-Term Predictions},
  author={Partha Ghosh and Jie Song and Emre Aksan and Otmar Hilliges},
  journal={2017 International Conference on 3D Vision (3DV)},
  year={2017},
  pages={458-466},
  url={https://api.semanticscholar.org/CorpusID:13549534}
}

@article{bib45,
  title={Visual ChatGPT: Talking, Drawing and Editing with Visual Foundation Models},
  author={Chenfei Wu and Sheng-Kai Yin and Weizhen Qi and Xiaodong Wang and Zecheng Tang and Nan Duan},
  journal={ArXiv},
  year={2023},
  volume={abs/2303.04671},
  url={https://api.semanticscholar.org/CorpusID:257404891}
}

@inproceedings{bib46,
  title={GLM: General Language Model Pretraining with Autoregressive Blank Infilling},
  author={Zhengxiao Du and Yujie Qian and Xiao Liu and Ming Ding and Jiezhong Qiu and Zhilin Yang and Jie Tang},
  booktitle={Annual Meeting of the Association for Computational Linguistics},
  year={2021},
  url={https://api.semanticscholar.org/CorpusID:247519241}
}

@inproceedings{bib47,
  title={librosa: Audio and Music Signal Analysis in Python},
  author={Brian McFee and Colin Raffel and Dawen Liang and Daniel P. W. Ellis and Matt McVicar and Eric Battenberg and Oriol Nieto},
  booktitle={SciPy},
  year={2015},
  url={https://api.semanticscholar.org/CorpusID:33504}
}

@inproceedings{bib49,
  title={GANs Trained by a Two Time-Scale Update Rule Converge to a Local Nash Equilibrium},
  author={Martin Heusel and Hubert Ramsauer and Thomas Unterthiner and Bernhard Nessler and Sepp Hochreiter},
  booktitle={Neural Information Processing Systems},
  year={2017},
  url={https://api.semanticscholar.org/CorpusID:326772}
}

@article{bib52,
  title={Low-rank multimodal fusion algorithm based on context modeling},
  author={Bai, Zongwen and Chen, Xiaohuan and Zhou, Meili and Yi, Tingting and Chien, Wei-Che},
  journal={Journal of Internet Technology},
  volume={22},
  number={4},
  pages={913--921},
  year={2021}
}

@article{bib58,
  title={Learning Individual Styles of Conversational Gesture},
  author={Shiry Ginosar and Amir Bar and Gefen Kohavi and Caroline Chan and Andrew Owens and Jitendra Malik},
  journal={2019 IEEE/CVF Conference on Computer Vision and Pattern Recognition (CVPR)},
  year={2019},
  pages={3492-3501},
  url={https://api.semanticscholar.org/CorpusID:182952539}
}

@article{bib61,
  title={An Image is Worth 16x16 Words: Transformers for Image Recognition at Scale},
  author={Alexey Dosovitskiy and Lucas Beyer and Alexander Kolesnikov and Dirk Weissenborn and Xiaohua Zhai and Thomas Unterthiner and Mostafa Dehghani and Matthias Minderer and Georg Heigold and Sylvain Gelly and Jakob Uszkoreit and Neil Houlsby},
  journal={ArXiv},
  year={2020},
  volume={abs/2010.11929},
  url={https://api.semanticscholar.org/CorpusID:225039882}
}

@article{bib62,
  title={Swin Transformer: Hierarchical Vision Transformer using Shifted Windows},
  author={Ze Liu and Yutong Lin and Yue Cao and Han Hu and Yixuan Wei and Zheng Zhang and Stephen Lin and Baining Guo},
  journal={2021 IEEE/CVF International Conference on Computer Vision (ICCV)},
  year={2021},
  pages={9992-10002},
  url={https://api.semanticscholar.org/CorpusID:232352874}
}

@article{bib64,
  title={Modeling Human Motion with Quaternion-Based Neural Networks},
  author={Dario Pavllo and Christoph Feichtenhofer and Michael Auli and David Grangier},
  journal={International Journal of Computer Vision},
  year={2019},
  volume={128},
  pages={855-872},
  url={https://api.semanticscholar.org/CorpusID:59158790}
}

@article{bib65,
  title={Quaternion-Valued Correlation Learning for Few-Shot Semantic Segmentation},
  author={Zewen Zheng and Guoheng Huang and Xiaochen Yuan and Chi-Man Pun and Hongzhi Liu and Wing-Kuen Ling},
  journal={IEEE Transactions on Circuits and Systems for Video Technology},
  year={2023},
  volume={33},
  pages={2102-2115},
  url={https://api.semanticscholar.org/CorpusID:253661872}
}

@article{bib66, 
  title={Style-based motion analysis for dance composition}, 
  author={Andreas Aristidou and Efstathios Stavrakis and Margarita Papaefthimiou and George Papagiannakis and Yiorgos Chrysanthou}, 
  journal={The Visual Computer}, 
  year={2018}, 
  volume={34}, 
  pages={1725-1737}, 
  url={https://api.semanticscholar.org/CorpusID:27531229}
}

@ARTICLE{bib67,
  author={Sheng, Bin and Li, Ping and Ali, Riaz and Chen, C. L. Philip},
  journal={IEEE Transactions on Cybernetics}, 
  title={Improving Video Temporal Consistency via Broad Learning System}, 
  year={2022},
  volume={52},
  number={7},
  pages={6662-6675},
  doi={10.1109/TCYB.2021.3079311}
}

@article{bib68,
  title={EAPT: Efficient Attention Pyramid Transformer for Image Processing},
  author={Xiao Lin and Shuzhou Sun and Wei Huang and Bin Sheng and Ping Li and David Dagan Feng},
  journal={IEEE Transactions on Multimedia},
  year={2021},
  volume={25},
  pages={50-61},
  url={https://api.semanticscholar.org/CorpusID:245536278}
}

@article{bib69,
  title={BaGFN: Broad Attentive Graph Fusion Network for High-Order Feature Interactions},
  author={Zhifeng Xie and Wenling Zhang and Bin Sheng and Ping Li and C. L. Philip Chen},
  journal={IEEE Transactions on Neural Networks and Learning Systems},
  year={2021},
  volume={34},
  pages={4499-4513},
  url={https://api.semanticscholar.org/CorpusID:238476689}
}

@InProceedings{Li_2023_ICCV,
    author    = {Li, Zinuo and Chen, Xuhang and Pun, Chi-Man and Cun, Xiaodong},
    title     = {High-Resolution Document Shadow Removal via A Large-Scale Real-World Dataset and A Frequency-Aware Shadow Erasing Net},
    booktitle = {Proceedings of the IEEE/CVF International Conference on Computer Vision (ICCV)},
    month     = {October},
    year      = {2023},
    pages     = {12449-12458}
}

@inproceedings{ijcai2023p129,
  title     = {A Large-Scale Film Style Dataset for Learning Multi-frequency Driven Film Enhancement},
  author    = {Li, Zinuo and Chen, Xuhang and Wang, Shuqiang and Pun, Chi-Man},
  booktitle = {Proceedings of the Thirty-Second International Joint Conference on
               Artificial Intelligence, {IJCAI-23}},
  publisher = {International Joint Conferences on Artificial Intelligence Organization},
  editor    = {Edith Elkind},
  pages     = {1160--1168},
  year      = {2023},
  month     = {8},
  note      = {Main Track},
  doi       = {10.24963/ijcai.2023/129},
  url       = {https://doi.org/10.24963/ijcai.2023/129},
}

@article{luo2023devignet,
  title={Devignet: High-Resolution Vignetting Removal via a Dual Aggregated Fusion Transformer With Adaptive Channel Expansion},
  author={Luo, Shenghong and Chen, Xuhang and Chen, Weiwen and Li, Zinuo and Wang, Shuqiang and Pun, Chi-Man},
  journal={arXiv preprint arXiv:2308.13739},
  year={2023}
}
\newpage

\vspace{11pt}
\vfill
\end{document}